\documentclass[aps,prX,amssymb,longbibliography,superscriptaddress,twocolumn]{revtex4-1}
\usepackage{graphicx}
\usepackage{bm,color}

\usepackage{amsmath}
\usepackage{bm,color}
\usepackage{multirow}
\usepackage{ulem}
\usepackage{xcolor}

\begin{document}

\title{
Multiple quantum scar states and emergent slow-thermalization in the flat-band system
}
\date{\today}

\author{Yoshihito Kuno}
\author{Tomonari Mizoguchi}
\author{Yasuhiro Hatsugai}
\affiliation{Department of Physics, University of Tsukuba, Tsukuba, Ibaraki 305-8571, Japan}

\begin{abstract}
Quantum many-body scars (QMBS) appear in a flat-band model with interactions on the saw-tooth lattice.
The flat-band model includes a compact support localized eigenstates, called compact localized state (CLS).
Some characteristic many-body states can be constructed from the CLSs at a low-filling on the flat-band. 
These many-body states are degenerate. 
Starting with such degenerate states we concretely show 
how to construct multiple QMBSs with different eigenenergies embedded in the entire spectrum. 
If the degeneracy is lifted by introducing hopping modulation or weak perturbations, 
these states lifted by these ways can be viewed as multiple QMBSs. 
In this work, we focus on the study of the perturbation-induced QMBS. 
Perturbed states, which are connected to the exact QMBSs in the unperturbed limit, 
indicate common properties of conventional QMBS systems, 
that is, a subspace with sub-volume or area law scaling entanglement entropy, which indicates the violation of the strong eigenstate thermalization hypothesis (ETH). Also for a specific initial state, slow-thermalization dynamics appears. We numerically demonstrate these subjects. 
The flat-band model with interactions is a characteristic example in non-integrable systems with the violation of the strong ETH and the QMBS.
\end{abstract}


\maketitle
\section{Introduction}
Localization phenomena have been the main topic in condensed matter \cite{AL50}.
While a wave function is spatially localized, from the modern point of view, the localization phenomena give a new insight to the fundamental questions in equilibrium statistical mechanics, that is, thermalization problem in isolated systems. 
In equilibrium statistical mechanics, even if a system is isolated, each eigenstate thermalizes even without coupling to heat bath \cite{Deutsch,D'Alessio,Srednicki,Rigol}. 
Every expectation values of local observables for every eigenstate corresponds 
to the values obtained by thermal ensemble (e.g. micro canonical ensemble). 
This prediction of the conventional statistical mechanics is called eigenstate thermalization hypothesis (ETH). 
In particular, if all eigenstates satisfy ETH, then it is called {\it strong ETH} \cite{Garrison}. 
For some typical condensed matter models, the strong ETH has been numerically verified \cite{Kim,Santos}.
However, the ETH is not universal. 
Localization phenomena give a counterexample, that is, it {\it does not thermalize}. 
In particular, breaking the ETH has been observed in the recent studies of the many-body localization (MBL), 
experimentally \cite{Choi2015,Schreiber,Lukin2019}. 
The origin of the breaking ETH comes from extensive numbers of emergent local integrals of motion (LIOM) \cite{Nandkishore,serbyn2020}.  
Hence, the extensive numbers of the LIOMs serves as local conserved quantities which makes the many-body system integrable from non-integrable. 
Then, as dynamical aspect of MBL, any non-entangled initial states do not thermalize. 
The theoretical study on this has been as a current trend in condensed matter physics \cite{Nandkishore,Abanin2019,Imbrie2017}. 

Furthermore, much recently, Rydberg atom simulators discover anomalous slow-thermalized dynamics \cite{Bernien,Ebadi2020,Bluvstein}. 
Some specific charge density patterns do show slow-thermalization behavior. 
The thermalization behavior depends on the choice of the initial state. Such slow- or non-thermalizations behavior do not appear for arbitrary initial states, which is essentially different from dynamical behavior of the MBL systems. 
This experimental observation indicates that the system is non-integrable as a whole, but there exist some atypical eigenstate (or subspace), where ETH is partially broken (strong ETH is broken) and these eigenstates have low entanglement entropy (EE).
These atypical eigenstates are currently called quantum many-body scar (QMBS). 
So far, motivated by the experimental discoveries, an effective spin model describing the Rydberg experimental system \cite{Bernien}, namely the PXP model, has been extensively studied \cite{Turner,Turner2,Choi,Ho2019}.  
By these pioneering theoretical studies, typical characters, or criteria, of systems with QMBS are listed as follows:
\begin{enumerate}
\renewcommand{\labelenumi}{(\roman{enumi})}
\item Some eigenstates in a many-body energy spectrum exhibit low EE, i.e., they obey the area law or the sub-volume law.
\item Most of the eigenstates in many-body system are thermal, that is, satisfy ETH. The system behaves as a non-integrable system as a whole. 
\item Quench dynamics for a set of specific initial states exhibits non- or slow-thermalization. 
Thermalization depends on initial states.
\end{enumerate}

Based on the pioneering theoretical studies of QMBS \cite{Turner,Turner2} and the above three {\it criteria} (i)-(iii), 
exploratory studies of the scar state have been conducted for some quantum spin models \cite{Moudgalya1,Moudgalya2,Schecter2019,Iadecola2019,Lin2019,Iadecola2020,McClarty,Wildeboer,Srivatsa,Shibata,Lee2020,Huang,Langlett,Zhao2021}, topological models \cite{Ok,Moudgalya2020,Jeyaretnam}, bosonic models \cite{Hudomal,Sinha,Zhao2020} and lattice gauge theoretical models \cite{Surace,Banerjee}. 
Also, $\eta$-pairing \cite{Yang,Zhang} has been revisited to explore a QMBS \cite{Vafek2017,Mark}. 

Although a few fermionic systems have been known to exhibit QMBS \cite{Yang,Zhang,DeTomasi2019,KMH2020} and a recent experiment of fermi-Hubburd model trapped in a tilted optical lattice has observed the presence of non-thermalized dynamics induced by a kinetic constraint \cite{Scherg}, 
what types of fermionic systems have QMBS has not been exhausted yet. 

In this work, a fermionic flat-band system in a one dimensional system is studied in detail. We show that under a suitable fine-tuning of hopping parameters, particle-filling, and interactions the flat-band model turns out to possess multiple QMBSs, which meet the three criteria (i)-(iii) mentioned above. We show the construction method of multiple QMBSs by extending the strategy of the QMBS in the previous work,
where a unique QMBS has been explicitly constructed by using the CLS \cite{KMH2020}. 
By employing hopping modulation or a weak perturbation, which induce energy splitting for degenerate many-body states obtained by CLSs, we construct the QMBS. 
In particular, we focus on the perturbative scheme, which is more realistic than the fine-tunig hopping scheme. In the perturbation scheme, the perturbed many-body states inherit the nature of the exact QMBSs and these perturbed many-body states span a small subspace decoupled other typical eigenstates, which are almost thermal and satisfy the ETH. 
Also, since the weakly perturbed many-body states remain to have low EE, which is an original property of the many-body states from CLSs, 
the flat-band model with weak perturbations exhibits a violation of the strong ETH \cite{Kim,Santos}. 
Furthermore, due to the presence of the multiple QMBS subspace, we can demonstrate slow-thermalization dynamics with fine-tuned entangled initial states. This is also a hallmark of the system with QMBS. 
Therefore, this subspace of the perturbed many-body states can be regarded as the set of QMBS.

So far, there are some works about perturbation effects to QMBS \cite{Choi,Khemani2020,Lin}. 
In particular, the work by Lin \cite{Lin} has focused on the fate of the QMBS in the PXP model under perturbations in detail. 
On the other hand, this work explores the opposite direction, that is, 
we make use of perturbation effects in that perturbation effects separate original degenerate many-body states obtained by CLSs, and then give multiple QMBS with different eigenenergies embedded in the thermal states.

This paper is organized as follows. In Sec.~II, we introduce the target flat-band model with interactions and discuss the CLSs of this system. Then we shows a constructions scheme of QMBS from a hopping modulation in Sec.~III. From Sec.~IV, we discuss the effects of an on-site linear potential 
and the effects of the perturbation are discussed in Sec.~VI. 
In Sec.~VII, we show the numerical demonstration of the presence of the QMBS. 
The spectrum structure and entanglement properties are investigated in detail. 
Finally, the dynamics and thermalization properties are numerically investigated.
Section VIII is devoted to the conclusion.

\section{Model}
\begin{figure}[t]
\centering
\includegraphics[width=8cm]{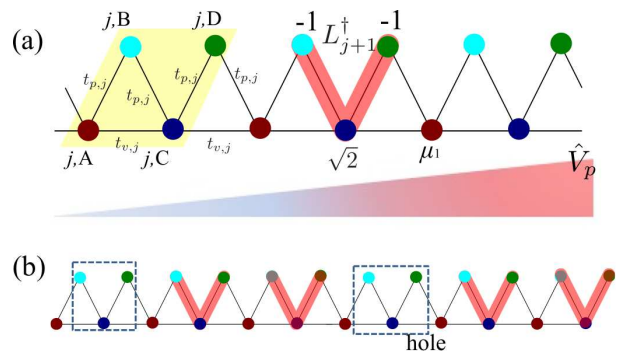}
\caption{(a) Schematic figure of the saw-tooth lattice model of Eq.~(\ref{Model}) with the linear potential ${\hat V}_{p}$.
The yellow shade regime is a unit cell. The V-shape red shade represents the CLS given by $L^{\dagger}_{j+1}$. 
(b) An example of the CLS many-body state given by Eq.~(\ref{Mscar}), $|S_{\ell}(N_{p}=4,L=6)\rangle$.}
\label{Fig1v}
\end{figure}

The target system is a spinless fermion system on the saw-tooth lattice described by the following Hamiltonian
\begin{eqnarray}
{\hat H}_{0}&=&\sum^{L-1}_{j=0}\biggl[
 t_{j,p}(f^{\dagger}_{j,A}f_{j,C}+f^{\dagger}_{j,C}f_{j+1,A}+\mbox{h.c.})\nonumber\\
&+&t_{j,v}(f^{\dagger}_{j,A}f_{j,B}+f^{\dagger}_{j,B}f_{j,C}
+f^{\dagger}_{j,C}f_{j,D}+f^{\dagger}_{j,D}f_{j+1,A}+\mbox{h.c.})\biggr]\nonumber\\
&+&\sum_{j,\alpha}\mu_{\alpha}f^{\dagger}_{j,\alpha}f_{j,\alpha},
\label{Model}
\end{eqnarray}
where $f^{(\dagger)}_{j,\alpha}$ ($\alpha=A,B,C,D$) is spinless fermion annihilation (creation) operators at site $(j,\alpha)$. 
$t_{j,p}$ and $t_{j,v}$ are a parallel and zig-zag hopping amplitudes, which depend on the unit cell $j$.  
$\mu_{\alpha}$ ($\alpha=A,B,C,D$) are on-site potentials. 
The schematic figure of the model is shown in Fig.~\ref{Fig1v}. We consider the $j$-dependent hopping case as $t_{j,p}=t_j$, $t_{j,v}=\sqrt{2}t_j$. 
Throughout this work, we focus on a finite system size $L$ and open boundary case.

The system has an orthogonal CLS
\begin{eqnarray}
L^{\dagger}_{j}=\frac{1}{2}\biggl[-f^{\dagger}_{j,B}+\sqrt{2}f^{\dagger}_{j,C}-f^{\dagger}_{j,D}\biggr].
\label{CLS}
\end{eqnarray}
The CLS $L^{\dagger}_j$ is also a creation operator of a fermion, i.e., regarded as a particle. 
Even for tuning of the hoppings, the CLSs are orthogonal to each other. 
If we set $\mu_B=\mu_C=\mu_D=\mu_1$ and $\mu_A=\mu_1+\mu_0$ ($\mu_1\neq 0$, $\mu_0<0$) \cite{remark1}, 
the single particle Hamiltonian is given by
\begin{eqnarray}
{\hat H}_0=\sum^{L-1}_{j=0}(\mu_0-2t_j)L^{\dagger}_{j}L_{j}+\sum_{\ell}\epsilon_{\ell}E^{\dagger}_{\ell}E_{\ell},
\end{eqnarray}
where $E^{(\dagger)}_{\ell}$ is an annihilation (creation) operator except for the CLSs, 
which is generally extended states, given by $E^{\dagger}_{\ell}=\sum_{j,\alpha}c^{\ell}_{j,\alpha}f^{\dagger}_{j,\alpha}$.
The Hamiltonian satisfies  $[L^{\dagger}_{j},{\hat H}_0]=(\mu_0-2t_j) L^{\dagger}_{j}$ and the CLS and extended states are orthogonal, $\{L^{\dagger}_j,E_k \}=0$ for any $j$ and $k$. 
For the unit-cell dependent hopping $t_j$, the single particle energies of each CLSs are be different. 
On the other hand, for a uniform hopping case, $t_j=\mu_0/2$ (uniform hopping amplitude) and $\mu_0<0 \:(>0)$ the zero-energy flat-band appears as the second (lowest) band; the other bands are dispersive. The existence of the single flat-band can also be dictated by the ``molecular-orbital" representation method \cite{Maruyama,Mizoguchi,Mizoguchi2021,Hatsugai2021}.

As our previous work \cite{KMH2020}, we consider a short-range interaction. 
Generally, while the interactions do not act between nearest-neighbor CLSs, 
the interaction makes the system non-integrable as a whole. 
Even for the Hamiltonian ${\hat H}_{0}$ with such short-range interactions, 
the CLS remains an (single particle) eigenstate. 
In this work, we set the following standard nearest-neighbor interaction,
\begin{eqnarray}
\hat{V}_{int}=V_{0}\sum_{j}&&\biggl[n^{A}_{j}n^{B}_{j}+n^{A}_{j}n^{C}_{j}+n^{B}_{j}n^{C}_{j}+n^{C}_{j}n^{D}_{j}\nonumber\\
&&+n^{C}_{j}n^{A}_{j+1}+n^{D}_{j}n^{A}_{j+1}\biggr],
\end{eqnarray}
where $V_0$ is the strength of the interaction. 
This interaction makes the system non-integrable as shown later.

Also, for the later discussion we introduce a linear potential 
\begin{eqnarray}
{\hat V}_{p}&=&\sum^{L-1}_{j=0}\mu_d\biggr[ 
(4j-i_0)f^{\dagger}_{j,A}f_{j,A}
+(4j-i_0+1)f^{\dagger}_{j,B}f_{j,B}\nonumber\\
&&+(4j-i_0+2)f^{\dagger}_{j,C}f_{j,C}
+(4j-i_0+3)f^{\dagger}_{j,A}f_{j,A}\biggl],\nonumber\\
\label{Vp}
\end{eqnarray}
where $i_0=(4L-1)/2$.
The introduction of the linear potential may be realistic for the implementation in coldatom experiments \cite{Meinert,Scherg}.

\section{Many-body state from CLS}
For the system ${\hat H}_0+\hat{V}_{int}$, we consider the $N_{p}$-particle system for $L$ unit-cell with the case $N_{p}\leq L$. 
Then, the following states are exact many-body eigenstates of the Hamiltonian: 
\begin{eqnarray}
|\Psi_{\rm L}(\{ j_k\}) \rangle =\prod^{N_p}_{k=1} L^{\dagger}_{j_k}|0\rangle,  
\label{Mscar}
\end{eqnarray} 
where $\{j_k\}$ ($k=1,2,\cdots, N_p$) corresponds to a set of $N_p$ numbers without duplication taken from a set of unit cell site $\{0,1,\cdots, L-1 \}$. We call the state $|\Psi_{\rm L}(\{ j_k\}) \rangle$ the {\it CLS many-body state}. 
If one sets $N_p=L$, corresponding to the fully occupied CLSs, then the choice of $\{j_k\}$ is unique and $|\Psi_{\rm L}(\{ j_k\}) \rangle$ can be regarded as a unique CLS many-body eigenstate discussed in \cite{KMH2020}. 

In this work, we focus on $N_p < L$ case, where 
the choice of $\{j_k\}$ is multiple, each of which is labeled by $\ell$ ($\ell=1, 2,\cdots, N_D$, $N_D={ L \choose Np}$). Then we denote each $|\Psi_{\rm L}(\{ j_k\}) \rangle$ by $|S_{\ell}(N_{p},L)\rangle$ for later discussions.
Each state $|S_{\ell}(N_{p},L)\rangle$ is an exact many-body eigenstate with energy $(\mu_0-2t_{j_k})$, $[{\hat H}_0+\hat{V}_{int}]|S_{\ell}(N_{p},L)\rangle=E_{\ell}(N_p,L)|S_{\ell}(N_{p},L)\rangle$, where $ E_{\ell}(N_p,L)=\sum^{N_p}_{k=1}(\mu_0-2t_{j_k})$. 
The schematic figure of the typical example of the CLS many-body state is shown in Fig.~\ref{Fig1v} (b). 
The interaction term ${\hat V}_{int}$ does not act on $|S_{\ell}(N_{p},L)\rangle$, i.e., zero eigenvalue, since the CLSs are spatially separated \cite{Hart,KMH2020}. 

For uniform hopping case $t_j=\mu_0/2$, 
the state $|S_{\ell}(N_{p},L)\rangle$ can be a kind of many-body state with ($L-N_p$)-holes on the flat-band, 
where each particles are the CLS, which is spatially localized with zero energy.

For later discussion, we here give the concrete definition of the EE. 
It is defined as the von-Neumann EE for a reduced density matrix for a subsystem, 
$S_{\rm e}=-{\rm Tr}\rho_A \ln \rho_A$, 
where $\rho_A={\rm Tr}_{B}|\Psi\rangle \langle \Psi|$ is a reduced density matrix, $|\Psi\rangle$ is a many-body eigenstate and system is divided into A and B subsystems.

\section{Construction of exact QMBS by fine-tuning of hoppings}
As one of the simplest ways to construct a characteristic system satisfying the criteria (i)-(ii) in Sec.~I, 
we fine-tune the distribution of the hopping amplitude $t_j$ in the system for the system ${\hat H}_0+\hat{V}_{int}$ (where ${\hat V}_p=0$).
The fine-tuning leads to the characteristic spectrum structure and entanglement properties. 
If one prepares the suitable set of $t_j$,  
each energy of $N_{D}$ many-body states $|S_{\ell}(N_{p},L)\rangle$, which we denote $E_{\ell}$, are not degenerate in principle, 
then the state $|S_{\ell}(N_{p},L)\rangle$ can be broadly embedded in the spectrum for the system ${\hat H}_0+\hat{V}_{int}$. As a simplest example, we can set $t_j=\beta(j-j_0)$, where $\beta$ is an arbitrary constant \cite{t_j_case2}. The $t_j$ acts as a tilted potential for the CLS. If we set the order of energy $\mathcal{O}(\beta)\sim \mathcal{O}(V_0)$, 
the state $|S_{\ell}(N_{p},L)\rangle$ can be broadly embedded in the spectrum (although not all degeneracy is necessarily lifted). Here, for large $V_0$, the system ${\hat H}_0+\hat{V}_{int}$ is non-integrable as a whole, most of eigenstates are thermal except for the state $|S_{\ell}(N_{p},L)\rangle$ with the eigenenergy $E_{\ell}(N_p,L)$. 
The distribution of the EE exhibits a characteristic distribution, i.e., 
the EEs for the $|S_{\ell}(N_{p},L)\rangle$ at the eigenenergy $E_{\ell}(N_p,L)$ in the spectrum exhibit low values compared to other thermal eigenstates. 
In particular, if one cuts the part of the system without the CLS in picking up the sub-system and measures the EE, the EEs for the $|S_{\ell}(N_{p},L)\rangle$ are exact zero (are-law EE) for any system size $L$.
Therefore, $N_{D}$ many-body states $|S_{\ell}(N_{p},L)\rangle$ can be regarded as QMBS. 
Although the exact QMBS states can be embedded in the system by this strategy, 
preparing the suitable distribution of $t_j$ is fairly artificial when assuming the implementation of the set of $t_j$ in real experimental systems such as coldatoms \cite{Zhang2015}. 
Therefore, we focus on another way to construct a characteristic system satisfying the criteria (i)-(iii) in Sec.~I from the next sections.


\section{A perturbation method by using the linear potential}

We show a more experimentally realizable setup to construct a characteristic system satisfying the criteria (i)-(iii).
To this end we hereafter focus on the uniform hopping case, $t_j=\mu_0/2$. Then, all CLS many-body states $|S_{\ell}(N_{p},L)\rangle$ are degenerate with zero-energy. 
Hence, a linear superposition of CLS many-body states $|S_{\ell}(N_{p},L)\rangle$ is also a certain eigenstate for ${\hat H}_{total}$. Such a state can have a large EE, at least not having area-law EE. The property of the EE for such a state are discussed in Appendix A.
Hence, in order to construct explicit multiple QMBSs in this system, it is better that such degeneracy is lifted as much as possible.  
To this end, we employ the finite linear potential ${\hat V}_{p}$, where $|\mu_{d}|\ll |t_{0}|, |V_{0}|$, that is, 
the linear potential ${\hat V}_{p}$ is perturbative \cite{tilt_scale}. 
Practically, the introduction of the linear potential ${\hat V}_p$ is more realistic than setting the fine-tuning of the unit-cell dependent hoppings $\{ t_j\}$. 
With finite ${\hat V}_{p}$, the CLS of Eq.~(\ref{CLS}) is {\it not} the single particle eigenstate for ${\hat H}_{0}+\hat{V}_p$, 
hence neither is the many-body state of Eq.~(\ref{Mscar}). 
If $\mu_d$ is large, 
the system can turn into the Wannier-Stark localization \cite{Schulz,Nieuwenburg,Orito,Kim,Taylor,Scherg}, but we do not focus on such a regime. 
For later purposes, we denote the total Hamiltonian by $\hat{H}_{tot}=\hat{H}_{0}+{\hat V}_{int}+{\hat V}_{p}$.


We discuss how to act the perturbation ${\hat V}_{p}$ for the degenerate CLS many-body states $|S_{\ell}(N_{p},L)\rangle$ (In what follows, we use the simpler notation, $|S_{\ell}\rangle$ where $\ell=1,\cdots, N_D$). 

Without ${\hat V}_p$, the states $|S_{\ell}\rangle$ are closed under swapping transformations between unit cells, that is, invariant for this swapping manipulation. ${\hat H}_0$ is also invariant for it. 
Here, if a suitably small value of $\mu_d$ in ${\hat V}_p$ is set, the $N_{D}$ degenerate states $|S_{\ell}\rangle$ are split in the first-order level, and  ${\hat V}_{p}$ weakly corrects the states $|S_{\ell}\rangle$. 
Practically, the energy splitting and the correction for the states $|S_{\ell}\rangle$ by ${\hat V}_p$ 
are quantitatively estimated by the degenerate perturbation theory. The states $|S_{\ell}\rangle$ are no longer {\it exact} eigenstates.
However, we expect that the corrected states are slightly different from the original states $|S_{\ell}\rangle$ and some physical properties of the state $|S_{\ell}\rangle$, such as particle distribution, entanglement, etc., are almost unchanged. 
From these expectations, 
many-body eigenstates generated by the correcting the CLS many-body states $|S_{\ell}\rangle$ can be viewed as QMBS. 

To demonstrate the above scenario, we first investigated two-particle system in detail. The results are given in Appendix B and C. 
The system is a minimum system exhibiting the character of the QMBS. 
The study of the two-particle system indeed gives an insight into larger systems. The two-particle system indicates the presence of the QMBS.

\begin{figure}[t]
\centering
\includegraphics[width=7cm]{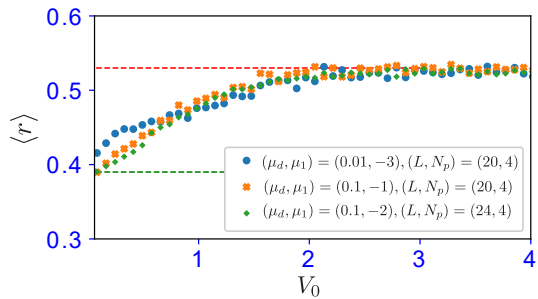}
\caption{
Mean level spacing ratio for the system with $L=5$, $N_p=4$ and for the system with $L=6$, $N_p=4$.
At $V_0=0.04$, $\langle r \rangle$ slightly deviates from $\sim 0.39$, due to degeneracy of energy eigenvalues.}
\label{Fig2v}
\end{figure}

\section{Numerical Demonstration}
In what follows, we numerically verify the expectation in Sec.~V by treating the system with numerically our accessible system size by exact diagonalization \cite{Quspin}.

\subsection{Level spacing analysis}
To begin with, we investigate the integrability of the system of ${\hat H}_{tot}$ by applying level spacing analysis \cite{Pal}. 
The integrability can be turned by the strength of interaction $\hat{V}_{int}$. 
Here, we set open boundary condition, diagonalize $\hat{H}_{total}$ directly, and obtain all energy eigenvalues.  
Then, we calculate the level spacing $r_s$ defined by $r_{s}=[{\rm min}(\delta^{(s)}, \delta^{(s+1)})]/[{\rm max}(\delta^{(s)},\delta^{(s+1)})]$ for all $s$, where $\delta^{(s)}=E_{s+1}-E_{s}$ and 
$\{E_{s}\}$ is the set of energy eigenvalue in ascending order 
and calculate the mean level spacing $\langle r\rangle$ which obtained by averaging over $r_{s}$ 
by employing all energy eigenvalues.
By varying $V_0$, the behavior of $\langle r\rangle$ is observed. 
When the system is integrable, the average level spacing takes $\langle r\rangle\simeq 0.39$, corresponding to the Poisson distribution. 
On the other hand, when the system is non-integrable, the average level spacing takes $\langle r\rangle\simeq 0.53$, corresponding to the Wigner-Dyson distribution \cite{Pal,Santos}.
Figure \ref{Fig2v} is the numerical result. 
As increasing $V_0$, $\langle r\rangle$ shows crossover from integrable to non-integrable. 
The result indicates that the interaction ${\hat V}_{int}$ makes the system non-integrable. 
\begin{figure}[t]
\centering
\includegraphics[width=9cm]{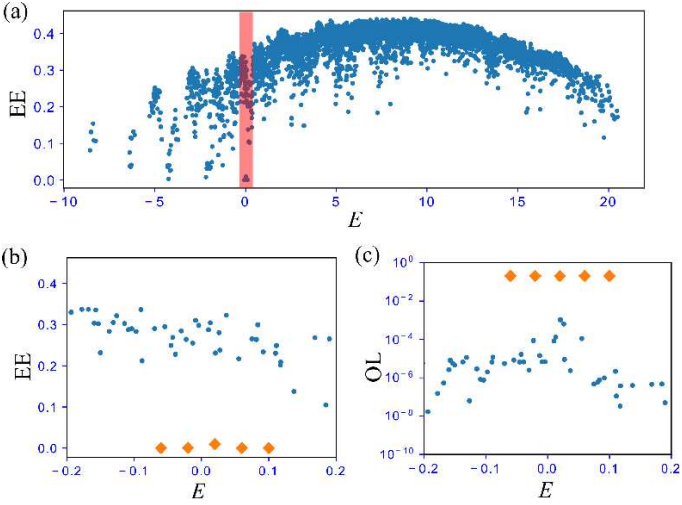}
\caption{
(a) Entire distribution of EE. The EE is normalized by the number of sites in the subsystem. 
(b) Zoom-up of distribution of the EE around $E=0$ ($E$ is many-body energy). 
The diamond labels represent the eigenstates with the large OL. The subsystem for the calculation of EE includes the lattice site, 
$(j,\alpha)=\{(0,A),(0,B),(0,C),(0,D),\cdots, ((L-3)/2,C),((L-3)/2,D)\}$ ($L$ is an odd integer).
(c) The distribution of OL for the eigenstates around $E=0$. 
We set $L=5$ with $N_p=4$ particles and $(\mu_d,V_0,\mu_1)=(0.01, 3,-3)$.
}
\label{Fig3}
\end{figure}
\subsection{Entanglement entropy and overlap for the detect state}
We investigate the EE for the system ${\hat H}_{tot}$.
Here, we should comment on the characteristic behavior about the EE of $|S_{\ell}\rangle$.
If one takes a subsystem where no CLS is cut in the calculation of the EE, the EE is zero. 
Trivially, even for thermodynamic limit, the EE of each $|S_{\ell}\rangle$ remains zero, corresponding to area-law of EE. 

Furthermore, to detect the presence of the corrected eigenstates coming from $N_D$ CLS many-body states $|S_{\ell}\rangle$ ($\ell=1,\cdots, N_D$), 
we introduce the following {\it detect state}, 
\begin{eqnarray}
|S^D_{(N_p,L)}\rangle =\sum^{N_{D}}_{\ell=1}\frac{1}{\sqrt{N_{D}}}|S_{\ell}\rangle.
\label{detect_state_0}
\end{eqnarray}
By using the state we measure an overlap, defined by 
\begin{eqnarray}
{\rm OL}=|\langle S^{D}_{(N_p,L)}|\psi_k\rangle|^2,
\label{OL}
\end{eqnarray}
where $|\psi_k\rangle$ is $k$-th eigenstate of the system of ${\hat H}_{tot}$. 
If $|\psi_k\rangle$ are close to a state $|S_{\ell}\rangle$. 
The order of the overlap is $\mathcal{O}(N_D^{-1})$, 
which is larger than that for other (thermal) states as far as finite size systems are concerned.

Since we showed the flat-band system turns into non-integrable system by the interaction ${\hat V}_{int}$, 
from now on, we ask if the system have QMBS within accessible system size. 
For the system without ${\hat V}_{p}$,  
the set of the CLS many-body states given by Eq.~(\ref{Mscar}) are exact eigenstates. 
However, once the perturbation ${\hat V}_p$ is switched on, the degeneracy is lifted as shown in the previous section. At the same time, the CLS many-body states are weakly perturbed. The corrected eigenstates tend to be much close to the individual CLS many-body states (not close to a linear superposed state the CLS many-body state). Thus, each exact eigenstate corrected from the CLS many-body states tends to be low entangled and can be viewed as QMBS.

\begin{figure}[t]
\centering
\includegraphics[width=9cm]{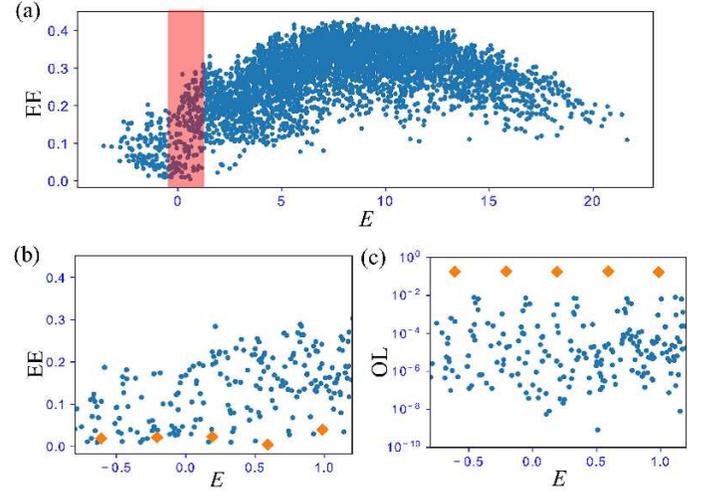}
\caption{
(a) Entire distribution of EE. The EE is normalized by the number of sites in the subsystem. (b) Zoom-up of the distribution of EE around $E=0$. The diamond labels represent the eigenstates with the large OL. The subsystem for the calculation of EE includes the lattice site, $(j,\alpha)=\{(0,A),(0,B),(0,C),(0,D),\cdots, ((L-3)/2,C),((L-3)/2,D)\}$ ($L$ is an odd integer).
(c) The distribution of OL for the eigenstates around $E = 0$.
We set $L=5$ with $N_p=4$ particles and $(\mu_d,V_0,\mu_1)=(0.1, 1,-1)$.}
\label{Fig4}
\end{figure}

We focus on the system with $L=5$, $N_p=4$ and $N_D=5$, where the five CLS many-body states are given by 
\begin{eqnarray}
&&|S_{1}\rangle = L^{\dagger}_{1}L^{\dagger}_{2}L^{\dagger}_{3}L^{\dagger}_{4}|0\rangle,\:\:
|S_{2}\rangle = L^{\dagger}_{0}L^{\dagger}_{2}L^{\dagger}_{3}L^{\dagger}_{4}|0\rangle,\nonumber\\
&&|S_{3}\rangle = L^{\dagger}_{0}L^{\dagger}_{1}L^{\dagger}_{3}L^{\dagger}_{4}|0\rangle,\:\:
|S_{4}\rangle = L^{\dagger}_{0}L^{\dagger}_{1}L^{\dagger}_{2}L^{\dagger}_{4}|0\rangle,\label{SD_ND=5}\\
&&|S_{5}\rangle = L^{\dagger}_{0}L^{\dagger}_{1}L^{\dagger}_{2}L^{\dagger}_{3}|0\rangle.\nonumber
\end{eqnarray}
The detect state $|S^{D}\rangle$ is also given by Eq.~(\ref{detect_state_0}) with $N_D=5$.

We first set the parameters, $(\mu_d,V_{0},\mu_1)=(0.01,3,-3)$ 
and calculate the EE for all eigenstates of the system as shown in Fig.~\ref{Fig3} (a). 
An arched distribution of the EE appears as a whole tendency. 
The highest value of EE at the spectrum center is roughly close to the maximum EE of the one-hole case, $S^{Max}_e/N_A=(2L-3)\ln 2/8\sim 0.6065$ where $N_A$ is the number of site of the subsystem \cite{Page_curve}. Hence, these results imply that most of the eigenstates are thermal and the system is non-integrable as a whole. 
We further focus on the EE around $E=0$ in the entire spectrum. 
There is a cluster of atypical states with low-valued EE. 
The zoom-up of the regime is shown in Fig.~\ref{Fig3} (b). 
There are five eigenstates with low valued EE. 
These states are decoupled from most of the typical eigenstates with large EE. These atypical states are close to the CLS many-body states. The finite energy splitting among these atypical states comes from the perturbation of ${\hat V}_{p}$.  
To clarify it we calculate the OL for the detect state $|S^{D}\rangle$. 
The results of the OL around $E=0$ are shown in Fig.~\ref{Fig3} (c). Certainly, five eigenstates corresponding to the five atypical eigenstates with low EE in Fig.~\ref{Fig3} (b) have large OL. 
Also, these atypical states exist in the thermal spectrum. Indeed, these atypical states can be viewed as QMBS in the sense that the criterion of (i) and (ii) as mentioned in Sec.~I are satisfied. 
This fact implies that our flat-band system exhibits violation of strong ETH \cite{Kim}. 
We also investigated the five QMBS from the degenerate perturbation theory [See Appendix E]. Even for the four-particle system, the degenerate perturbation theory captures the QMBS close to the CLS many-body states.
We will further show dynamics generated from the subspace of QMBS later. 

In the same system, we further study large $\mu_d$, small $V_0$ and small $|\mu_1|$ case, $(\mu_d,V_0,\mu_1)=(0.1,1,-1)$, where the perturbation picture becomes ambiguous and small interactions make non-integrability weak. 
Figure \ref{Fig4} (a) is a whole distribution of the EE for all eigenstates. 
The distribution indicates a wider range of EE values than that in Fig.~\ref{Fig3} (a). 
This implies that the integrability is weak. 
Under this situation, in Fig.~\ref{Fig4} (b) we plot zoom-up of the distribution of EE around $E=0$. 
We again find five low entangled states with OL for $|S^D\rangle$. 
The OL is shown in Fig.~\ref{Fig4} (c). 
However, these low-entangled states are not isolated, 
and other eigenstates with low valued EE exist. 
Even for some large $\mu_d$, however, the five exact eigenstates close to the CLS many-body state $|S_{\ell}\rangle$ exist. 

\begin{figure}[t]
\centering
\includegraphics[width=9cm]{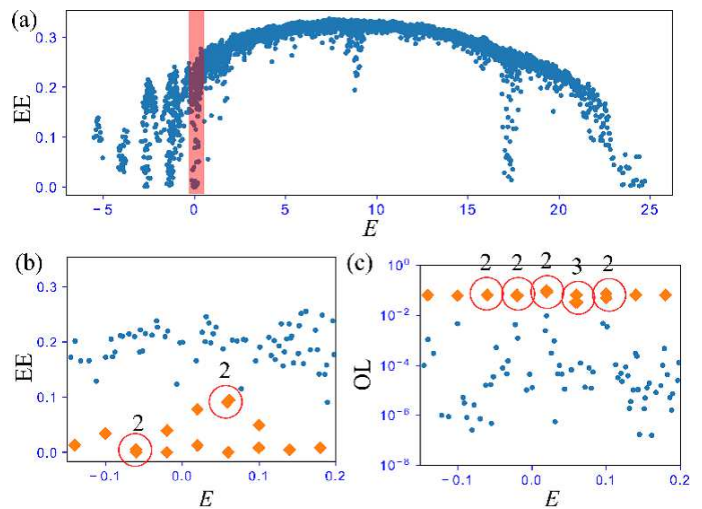}
\caption{
(a) Entire distribution of EE. The EE is normalized by the number of sites in the subsystem. 
(b) Zoom-up of the EE around $E=0$. The diamond labels represent the atypical eigenstates. 
(c) The distribution of OL for the eigenstates around $E = 0$.
The subsystem for the calculation of EE includes the lattice site, $(j,\alpha)=\{(0,A),(0,B),(0,C),(0,D),\cdots, ((L-2)/2,C),((L-2)/2,D)\}$.
We set $L=6$ with $N_p=4$ particles and $(\mu_d,V_0,\mu_1)=(0.1, 4,-2)$.}
\label{Fig5}
\end{figure}

In addition, we also consider the case of the larger system size and fewer particles, i.e., the system with $L=6$, $N_p=4$, 
with $(\mu_d,V_0,\mu_1)=(0.1, 4,-2)$.
For this case, there are $N_{D}=15$ CLS many-body states $|S_{\ell}\rangle$. 
As in the previous calculations, we calculate a whole distribution of EE as shown in Fig.~\ref{Fig5} (a). 
The highest value of EE at the spectrum center is roughly close to the maximum EE of two-hole case, $S^{Max}_e/N_A=(2^{N_A-1}-{N_A \choose N_A/2}-{N_A \choose N_A/2-1})/N_A\sim 0.4838$ where $N_A=10$. 
Hence, these results imply that most of the eigenstates are thermal and the system is non-integrable as a whole. 
Even for this case we also find a cluster of low entangled states around $E=0$. 
The zoom-up around $E=0$ is shown in Fig.~\ref{Fig5} (b). 
We find $15$ atypical states with low-valued EE. Some atypical states are degenerate since the linear potential ${\hat V}_{p}$ does not resolve all CLS many-body states $|S_{\ell}\rangle$. 
As shown in Fig.~\ref{Fig5} (c), these atypical states also have large OL for the detect state $|S^{D}\rangle$ even if they are partially degenerate.  
Even for the presence of degeneracy, under this parameter set, the EEs of the $15$ atypical states with large OL are small as a whole. This also implies the violation of strong ETH although some degenerate states do not exhibit area-law EE and the scaling law of the EE is logarithmic, $\mathcal{O}(\ln L)$ [See Appendix A]. 

\begin{figure}[t]
\centering
\includegraphics[width=8.5cm]{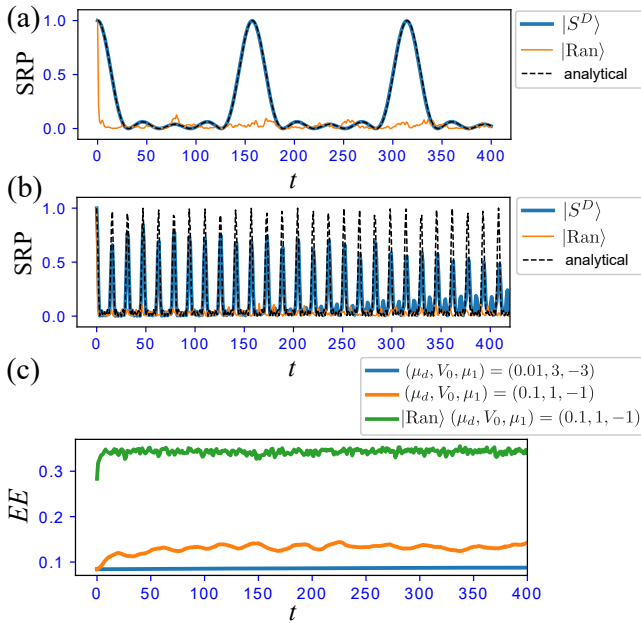}
\caption{(a) Dynamics of the SRP for $(\mu_d,V_0,\mu_1)=(0.01, 3,-3)$.
(b) Dynamics of the SRP for $(\mu_d,V_0,\mu_1)=(0.1,1,-1)$.
The black dashed line represents the analytical solution obtained in Appendx D.
(c) Dynamics of EE for $(\mu_d,V_0,\mu_1)=(0.01, 3,-3)$ and $(0.1, 1,-1)$. 
For $(\mu_d,V_0,\mu_1)=(0.1, 1,-1)$ case, the results for two initial states $|S^{D}\rangle$ and $|{\rm Ran}\rangle$ exhibit clear difference.
The subsystem for the calculation of EE includes the lattice site, $(j,\alpha)=\{(0,A),(0,B),(0,C),(0,D),\cdots, ((L-2)/2,C),((L-2)/2,D)\}$ and the EE is normalized by the number of lattice sites of the subsystem.
For all results, we set $L=5$ with $N_p=4$.}
\label{Fig6}
\end{figure}
\subsection{Slow-thermalization}
In the previous section, we show the presence of low entangled eigenstates with large overlap for the detect state $|S^{D}\rangle$. These states can be viewed as QMBS. 
To firmly characterize the presence of the QMBS in our flat-band system, 
in this section, we demonstrate slow-thermalization. 
The two-particle case clearly exhibits such slow-thermalization as shown in Appendix B.
To demonstrate it numerically, we calculate the squared return probability (SRP),
\begin{eqnarray}
({\rm SRP})=|\langle \Psi_0|\Psi(t)\rangle|^{2},
\label{SRP}
\end{eqnarray}
where $|\Psi_0\rangle$ is an initial state and $|\Psi(t)\rangle=e^{-i{\hat H}_{tot}t}| \Psi_0\rangle$. 
Here let us set the detect state $|S^{D}\rangle$ to the initial state 
and calculate the dynamics of the SRP.

Here, we focus on the system with $L=5$, $N_p=4$ and set $(\mu_d,V_{0},\mu_1)=(0.01,3,-3)$ for $\hat{H}_{tot}$. 
In the measurement of the dynamics, we employ the detect state $|S^{D}\rangle$ as an initial state,  
\begin{eqnarray}
|S_{ini}\rangle =|S^D_{(4,5)}\rangle=\sum^{5}_{\ell=1}\frac{1}{\sqrt{5}}|S_{\ell}\rangle,
\label{initial}
\end{eqnarray}
where $|S_{\ell}\rangle$ is the five CLS many-body state defined by Eq.~(\ref{SD_ND=5}) \cite{exp}. 
We observe the quench dynamics of the system ${\hat H}_{tot}$ for the initial state $|S_{ini}\rangle$. 
The strategy of choosing such an entangled state as an initial state for slow- or non-thermalized dynamics is similar to that in \cite{Iadecola2020}, where an entangled initial state constructed by magnon excited states was considered.

The result for the SPR is given in Fig.~\ref{Fig6}. 
The SPR exhibits almost complete revival behavior. 
The revival time is long due to the small energy splitting for the QMBS states as shown in Fig.~\ref{Fig3} while the SRP for a random state suddenly decays. 
The dynamics for $|S_{ini}\rangle$ is almost governed by the subspace of the QMBS. The initial state of $|S_{ini}\rangle$ exhibits slow-thermalization. 

For larger $\mu_d$ and small $V_0$ case $(\mu_d,V_{0},\mu_1)=(0.1,1,-1)$ as in Fig.~\ref{Fig4}, the behavior of the SRP is shown in Fig.~\ref{Fig6} (b). 
The SRP for $|S_{ini}\rangle$ gradually decays for a long time. 
Here, the subspace of eigenstates with the large OL for $|S^{D}\rangle$ is not completely decoupled from the other thermally eigenstates due to strong perturbation of ${\hat V}_{p}$ as indicated by the distribution of EE in Fig.~\ref{Fig4} (b). As shown in Fig.~\ref{Fig4} (c), the other eigenstates also have finite OL larger than those in Fig.~\ref{Fig3} (c). These factors induce gradual-decay. Needless to say, even though the eigenstates with the large OL for $|S^{D}\rangle$ in the case $(\mu_d,V_{0},\mu_1)=(0.1,1,-1)$ are not exact QMBSs, there is a clear difference in the dynamics, that is, slow-thermalization appears depending on the choice of the initial state. This means that the criterion (iii) given in Sec.I is satisfied in the flat-band model ${\hat H}_{tot}$.

We also calculate the dynamics of EE for both parameter sets as shown in Fig.~\ref{Fig6} (c) 
The EE for both parameters keeps low value in time evolution. 
However, there is a slight increase of the EE for the case $(\mu_d,V_{0},\mu_1)=(0.1,1,-1)$ in parallel with the gradual decaying behavior of the SRP. Compared to a random initial state, which exhibits sudden thermalization, the EE increase is much small as shown in Fig.~\ref{Fig6} (d).

Finally, we comment on the eventual behavior of the dynamics. 
We expect that the SRP for both cases approach zero eventually. Such a decaying behavior also has been expected in the bare PXP model studied in \cite{Turner,Turner2,Choi} and the decay can be also related to the proximity to an integrable point suggested in \cite{Khemani}. 
In the flat-band system, the study of dynamics for further large system size is interesting, and what the eventual behavior is remains an open question.

\section{Conclusion}
In this work, we have investigated the presence of multiple QMBSs 
in a flat-band system with interactions. 
Especially, we focused on the effects of a weak perturbation in the system.  
The main strategy to implement QMBSs is setting a fractional filling so that degenerate CLS many-body states are prepared. 
Then by introducing weak perturbations the degenerate states are split with different eigenenergies. 
The split many-body states remain low entangled property compared to typical thermal eigenstates 
and form a subspace where the EE does not at least obey the volume law. 
Furthermore, the presence of the subspace induces slow-thermalization only for a specific initial state. 
From the nature of the static spectrum, low-valued EEs, and the emergence of non-thermalized dynamics, 
the many-body eigenstates corrected from the CLS many-body states by weak perturbations can be viewed as QMBS. 
We numerically demonstrated the presence of QMBS within numerically accessible system size by exact diagonalization. 
The numerical results indicated that the three criteria (i)-(iii) given in Sec.I are satisfied in our flat-band system.

From these results, it is expected that the flat-band model in which the dispersive band and the flat-band coexist is one of the useful platforms for implementing QMBS. 

Needless to say, for a finite-size system, the slow-thermalization in the flat-band system can be controlled, if we set a suitable parameter set as in Figs.~\ref{Fig6} (a) and \ref{Fig6} (b). If its dynamics is realized in real experimental systems, its dynamics exhibits almost no-thermalization. 

We further comment that this perturbation scheme can be applicable for  flat-band systems in any dimensions and also, in this work, though we consider a linear potential of $\hat{V}_p$ [Eq.~(\ref{Vp})] for simplicity. Generic perturbations are also possible to lead same findings in our work.

\section*{Acknowledgments}
The work is supported in part by JSPS KAKENHI Grant Numbers JP17H06138 (Y.K, Y.H.), JP21K13849 (Y.K.) and JP20K14371 (T.M.). 

\renewcommand{\thesection}{A\arabic{section}} 
\renewcommand{\theequation}{A\arabic{equation}}
\setcounter{equation}{0}

\section*{Appendix A: Property of entanglement entropy of many-body states given by the set of $|S_{\ell}\rangle$}

In this appendix, we show the property of the EE of the degenerate bases, $|S_{\ell}\rangle$. 
Since all $|S_{\ell}\rangle$ are orthogonal to each other, 
the property of the EE of the state obtained from the set of $|S_{\ell}\rangle$ can be captured by the following mixed state density matrix with equal weight
\begin{eqnarray}
\rho^{M}=\sum^{N_D}_{\ell=1}|S_{\ell}\rangle \langle S_{\ell}|.
\label{rho_M}
\end{eqnarray}
Here, We assume $N_p (\geq L/2)$ particle system and a subsystem with system size $L_A=L/2$ where no CLS is cut. 
Since for this cut the state $|S_{\ell}\rangle$ has zero EE, 
the Schmidt decomposition for $|S_{\ell}\rangle$ is a simple form with single singular value, 
\begin{eqnarray}
|S_{\ell}\rangle = |S^{A}_{\ell}\rangle \otimes |S^B_{\ell}\rangle.
\label{detect_state}
\end{eqnarray}
From this fact, the partial density matrix $\rho^{M}_A$ is directly obtained, 
\begin{eqnarray}
\rho^{M}_A&=&{\rm Tr}_B  \rho^{M}
=\bigoplus_{k=0}^{L-N_{p}}\rho^{d}_k,\\
\rho^{d}_k&=& \frac{N^{B}_k}{L} {\bf 1}_{N^{A}_{k}},
\label{rho_A}
\end{eqnarray}
where $N^{A}_{k}={L/2 \choose L-N_p-k}$, $N^{B}_{k}={L/2 \choose k}$ and ${\bf 1}_{N^{A}_{k}}$ is $N^{A}_k\times N^{A}_{k}$ identity matrix.
From the form of $\rho^{M}_A$, the EE denoted by $S^{M}_e$ is given by 
\begin{eqnarray}
S^{M}_{e}=\sum^{L-N_{p}}_{k=0}\biggl[-\frac{N^{A}_{k}N^{B}_{k}}{L}\ln \frac{N^{B}_{k}}{L}\biggr].
\label{upper}
\end{eqnarray}
This EE gives the character of the system size dependence of a many-body state obtained from the set of $|S_{\ell}\rangle$.

In particular, let us consider $N_p=L-1$ case. Then the EE is explicitly given by 
\begin{eqnarray}
S^{M}_{e}=\frac{1}{2}\ln L +\ln 2.
\end{eqnarray}
The EE does not obey area-law. 
This result implies that many-body states composed of the superposition of $|S_{\ell}\rangle$ can have the EE with the order $\mathcal{O}(\ln L)$, not obeying area-law.

\section*{Appendix B: Study of two particle system}
In this Appendix, we show that even for two-particle system the system of ${\hat H}_{tot}$ exhibits the tendency of the presence of multiple QMBSs, that is, satisfies the three criteria mentioned in Sec.I. The detailed investigation of the two-particle system gives an insight into larger many-particle systems. 
In what follows, we consider $L=3$ and $N_p=2$ system with open boundary condition (OBC). 
This situation corresponds to one hole case in the flat-band.   
Then, in non-interacting system ${\hat H}_{0}$, 
$N_D=3$ two-particle states constructed by Eq.~(\ref{Mscar}) are   
\begin{eqnarray}
|S_{1}\rangle = L^{\dagger}_{0}L^{\dagger}_{1}|0\rangle,\:\:
|S_{2}\rangle = L^{\dagger}_{1}L^{\dagger}_{2}|0\rangle,\:\:
|S_{3}\rangle = L^{\dagger}_{0}L^{\dagger}_{2}|0\rangle.\nonumber\\
\label{Sells_two}
\end{eqnarray}
Without ${\hat V}_{p}$, all states $|S_{\ell}\rangle$ are exact eigenstates and degenerate with zero energy.

Under the perturbation, ${\hat V}_{p}$, the original two-particle CLS many-body states $|S_{\ell}\rangle$ are slightly corrected and embedded in the eigenenergy of the two-particle system. 
Concretely, we can estimate the perturbation effects by the degenerate perturbation theory, which is shown in Appendix C.

\begin{figure}[t]
\centering
\includegraphics[width=8.5cm]{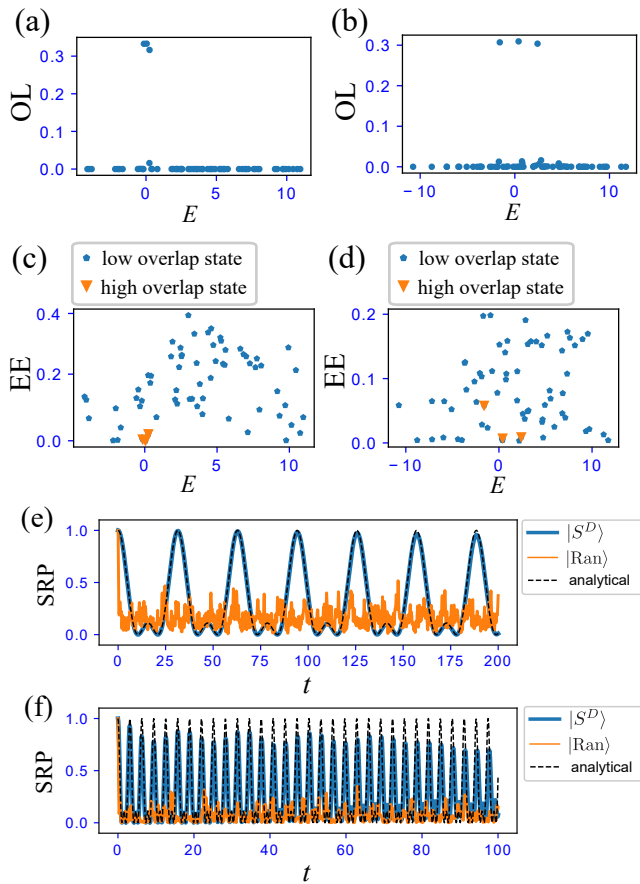}
\caption{Numerical results of two-particle system; 
(a) Overlaps for $(\mu_d,V_0,\mu_1)=(0.05, 5,-3)$. (b) Overlaps for $(\mu_d,V_0,\mu_1)=(0.5, 1,-5)$.
(c) Distribution of EE for $(\mu_d,V_0,\mu_1)=(0.05, 5,-3)$. (d) Distribution of EEs for $(\mu_d,V_0,\mu_1)=(0.5, 1,-5)$.
The EE is normalized by the number of sites in the subsystem. 
In the results of (c) and (d), the subsystem for the calculation of EE includes the lattice site, $(j,\alpha)=\{(0,A),(0,B),(0,C),(0,D)\}$. 
Dynamics of RP for $(\mu_d,V_0,\mu_1)=(0.05, 5,-3)$ (e), $(\mu_d,V_0,\mu_1)=(0.5, 1,-5)$ (f). The black dashed line represents the analytical solution obtained in Appendx D.}
\label{Fig7}
\end{figure}

Numerically, for the two-particle system of ${\hat H}_{tot}$, the OL given by Eq.~(\ref{OL}) is calculated as shown in Fig.~\ref{Fig7} (a) and (b). 
We set $(\mu_d,V_{0},\mu_1)=(0.05,5,-3)$ and $(\mu_d,V_{0},\mu_1)=(0.5,1,-5)$. 
In $(\mu_d,V_0,\mu_1)=(0.05, 5,-3)$ case, 
there are three eigenstates with large overlap. 
These states have different energies from each other around zero energy. 
The energy splitting of these states for finite $\mu_d$ comes from the perturbation effect of ${\hat V}_p$. 
These three states merge at zero energy in the limit ${\hat V}_{p}\to 0$. 
These states with large overlap are close to $|S_{\ell}\rangle$. 
Figure~\ref{Fig7} (c) shows the distribution EE. 
There, we see the three low EE states (inverted triangle label) around zero energy. 
From these results, we find that the weak perturbation ${\hat V}_{p}$ does not change the state $|S_{\ell}\rangle$ significantly and these states tend to have low EE compared to other extended eigenstates. 
These facts can be verified by the degenerate perturbation theory, shown in Appendix C. 

Furthermore, the OL for larger $\mu_d$ case is shown in Fig.~\ref{Fig7} (b), where we set $(\mu_d,V_0,\mu_1)=(0.5, 1,-5)$. 
Although the perturbation picture becomes subtle, there are three eigenstates with large overlap, which can be traced back to the degenerate states $|S_{\ell}\rangle$. 
The splitting of energies is larger than that in Fig.~\ref{Fig7} (a). The EE distribution is also shown in Fig.~\ref{Fig7} (d). The EE for the exact eigenstates close to the state $|S_{\ell}\rangle$ are small, but they are not separated from other states.  
The reason is that (I) large $\mu_d$ strongly perturbs the state $|S_{\ell}\rangle$, that is, mix other thermal-like eigenstates to enhance entanglement, (II) due to the small interaction the integrable tendency remains to leads small EEs as a whole. 
However, there certainly exist three eigenstates close to the state $|S_{\ell}\rangle$ with low-valued EE. 
Therefore, we also call the three eigenstates with large overlap QMBS.

From the character of eigenstates of the two-particle system, we can demonstrate slow-thermalized dynamics by setting the detect state $|S^{D}\rangle$ as an initial state.
The numerical results of the dynamics of the SRP are shown in Fig.~\ref{Fig7} (e) and (f). 
For $(\mu_d,V_0,\mu_1)=(0.05, 5,-3)$ case in Fig.~\ref{Fig7} (e), 
the SRP for the initial state $|S^{D}\rangle$ oscillates with large period since the energy splitting of the QMBS is small due to the small $\mu_d$, but the very small decay exists. 

The dynamics of the SRP include two frequencies understood by a simple analytical calculation in Appendix D. 
The analytical solution is plotted in Fig.~\ref{Fig7} (e) and the solution almost matches the numerical result. 
The result indicates slow-thermalization. 
On the other hand, for a random initial state $|{\rm Ran}\rangle$, 
the SRP decays suddenly. Thus, thermalizing property depends on the choice of the initial state. 

The result for larger $\mu_d$ and small interaction case, $(\mu_d,V_0,\mu_1)=(0.5, 1,-5)$ is shown in Fig~\ref{Fig7} (f), where the perturbation ${\hat V}_p$ works strongly. While the SRP for a random state $|{\rm Ram}\rangle$ decays suddenly, the SRP for the initial state $|S^{D}\rangle$ exhibits slow-thermalization. 
The oscillation period of the SRP is smaller than that in Fig.~\ref{Fig7} (e) since the energy differences of the QMBS are large. The amplitude of the SPR for the initial state $|S^{D}\rangle$ gradually decays since the QMBS coming from $|S_{\ell}\rangle$ is slightly hybridized with other extended eigenstates although the QMBS have a large overlap for $|S^{D}\rangle$. 
Compared with the analytical solution obtained from Appendix D, the numerical result deviates from the analytical one during the time evolution.

From these numerical results, the two-particle system exhibits the characteristic properties, the presence of QMBS with low EE and slow-thermalization for a specific initial state, related to the criteria (i) and (iii) in Sec.~I.

\begin{figure*}[t]
\centering
\includegraphics[width=18cm]{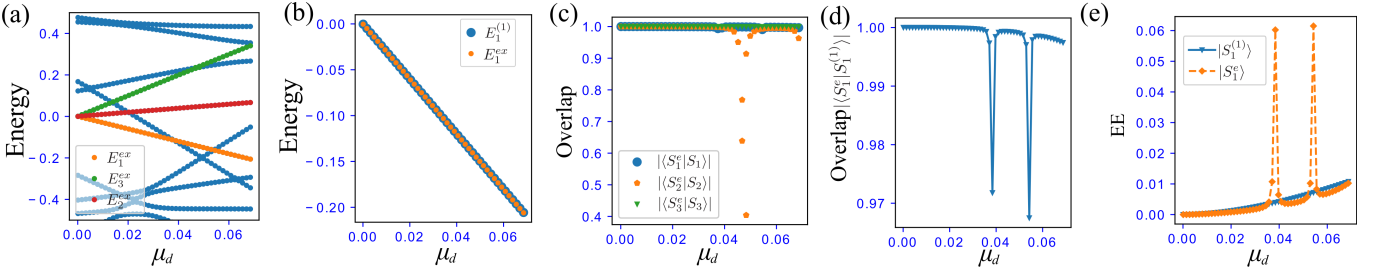}
\caption{(a) Behavior of energy eigenvalues as a function of $\mu_d$.  
(b) Comparison of exact energy of the eigenstate 
$|S^{e}_{1}\rangle$ with large OL with $|S_{1}\rangle$ 
and the first-order corrected energy for $|S_{1}\rangle$. 
(c) Overlap between the state $|S^{e}_{\ell}\rangle$ and the state $|S_{\ell}\rangle$. 
Even with the increase of $\mu_d$, large overlap appears.
(d) Overlap between the eigenstate $|S^{e}_{1}\rangle$ and the the first-order corrected state $|S^{(1)}_{1}\rangle$.
(e) Comparison of the EE for the eigenstate $|S^{e}_{1}\rangle$ and the EE of the first-order corrected state $|S^{(1)}_{1}\rangle$.
For all data, we set $L=3$ with $N_p=2$ particles, $(V_0,\mu_1)=(5,-3)$.}
\label{Fig8}
\end{figure*}

\section*{Appendix C: Degenerate perturbation theory for two particle system}

Without the perturbation ${\hat V}_{p}$, 
the three CLS many-body states given by Eq.~(\ref{Sells_two}), $|S_\ell\rangle$ ($\ell=1,2,3$) are degenerate. Actual eigenstates in numerics are generally given as linear superposed states from the three states $|S_{\ell} \rangle$. 

For small $\mu_d$ in ${\hat V}_{p}$, we employ the degenerate perturbation theory for the three states $|S_\ell\rangle$ 
and observe how much the three states $|S_\ell\rangle$ are affected by $\hat{V}_p$. 

In the degenerate perturbation theory, if one takes the orthogonal states $|S_{1}\rangle$, $|S_{2}\rangle$, and $|S_{3}\rangle$ in Eq.~(\ref{Sells_two}) as non-perturbative states, the secular equation for the first-order energy shift becomes just diagonal,
\widetext
\begin{eqnarray}
\left[
    \begin{array}{ccc}
      \langle S_1|\hat{V}_{p}|S_{1}\rangle -E^{(1)}_1 & 0 & 0 \\
      0 & \langle S_2|\hat{V}_{p}|S_{2}\rangle -E^{(1)}_2 & 0 \\
      0 & 0 & \langle S_3|\hat{V}_{p}|S_{3}\rangle-E^{(1)}_3
    \end{array}
  \right]
  \left[
  \begin{array}{c}
    c^0_1\\
    c^0_2\\
    c^0_3
   \end{array}
\right]
=0,
\end{eqnarray}
\endwidetext
\noindent 
where $c^{0}_j$ is coefficients to determine zero-th order states, $E^{(1)}_{\ell}$ is the first-order energy shift and these energies $E^{(1)}_{\ell}$ are non-degenerate. Also, the diagonal elements $\langle S_\ell|\hat{V}_{p}|S_{\ell}\rangle$ are real and the off-diagonal matrix elements become $\langle S_k|\hat{V}_{p}|S_{j}\rangle=0$ for $k\neq j$ since ${\hat V}_p$ does not transfer particles and the spatial particle configuration of the CLSs leads to zero overlap. 
If the values of each diagonal elements $\langle S_\ell|\hat{V}_{p}|S_{\ell}\rangle$ are different, 
the three degenerate states are split in the first-order level. This splitting depends on the form of ${\hat V}_p$, system size, and particle numbers, etc.  
There can be a care where the degeneracy cannot be lifted in the first-order level. 
In zero-th order level, the corrected states are just the CLS many-body states $|S_{\ell}\rangle$ ($\ell=1,2,3$) with different energies $0+E^{(1)}_{\ell}$. Beyond the zero-th order level, the first-order corrected states are given by 
\begin{eqnarray}
|S^{(1)}_{\ell}\rangle &=& |S_{\ell}\rangle + \sum_{\ell\:(\ell'\neq \ell)}A^{(1)}_{\ell,\ell'}|S_{\ell'}\rangle
+\sum_{k}B^{(1)}_{k,\ell}|k^{(0)}\rangle,
\label{1stPstate}\\
A^{(1)}_{\ell,\ell'}&=&\frac{\sum_{k}B^{(1)}_{k,\ell} 
\langle k^{(0)}|{\hat V}_{p}|S_{\ell'}\rangle}{E^{(1)}_{\ell}-E^{(1)}_{\ell'}},\\
B^{(1)}_{k,\ell}&=&\frac{\langle k^{(0)}|{\hat V}_{p}|S_{\ell}\rangle}{E^{(0)}_{\ell}-E^{(0)}_{k}}, 
\end{eqnarray}
where $|k^{(0)}\rangle$ is an eigenstate of the non-perturbed Hamiltonian $\hat{H}_0+{\hat V}_{int}$ except for $|S_{\ell}\rangle$. 
The above first-order corrected states $|S^{(1)}_{\ell}\rangle$ can be numerically calculated since the state $|S^{(1)}_{\ell}\rangle$ is invariant for a gauge transformation $|k^{(0)}\rangle \to \pm |k^{(0)}\rangle$. 
This condition is enough to proceed the numerical calculation because $|k^{(0)}\rangle$ is real vector from the real symmetric Hamiltonian matrix $\hat{H}_{0}+\hat{V}_{int}$, that is, gauge indefiniteness is only $\pm |k^{(0)}\rangle$.   

\begin{figure*}[t]
\centering
\includegraphics[width=18cm]{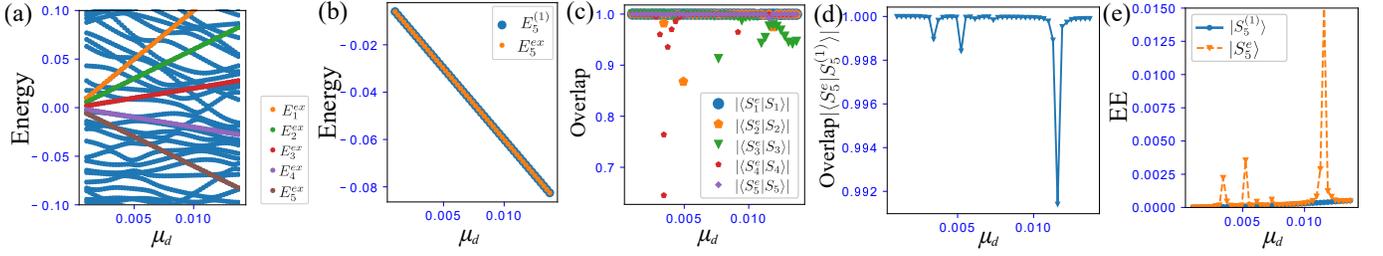}
\caption{(a) Behavior of energy eigenvalues as a function of $\mu_d$. 
(b) Comparison of exact energy of the eigenstate 
$|S^{e}_{5}\rangle$ with large OL with $|S_{5}\rangle$ 
and the first-order corrected energy for $|S_{5}\rangle$. 
(c) Overlap between the state $|S^{e}_{\ell}\rangle$ and the state $|S_{\ell}\rangle$. 
Even with the increase of $\mu_d$, large overlap appears.
(d) Overlap between the eigenstate $|S^{e}_{5}\rangle$ and the the first-order corrected state $|S^{(1)}_{5}\rangle$.
(e) Comparison of the EE for the eigenstate $|S^{e}_{5}\rangle$ and the EE of the first-order corrected state $|S^{(1)}_{5}\rangle$. 
For all data, we set $L=5$ with $N_p=4$ particles, $(V_0,\mu_1)=(3,-3)$.}
\label{Fig9}
\end{figure*}

For the two-particle system shown in Appendix B, 
we carried out the numerical calculation 
as varying $\mu_d: 0\to 0.07$. 
Figure~\ref{Fig8} (a) is the flow of the exact spectrum around $E=0$ with increase of $\mu_d$ by exact diagonalization. 
Three spectral lines extending from $E = 0$ at $\mu_d=0$ denoted by $E^{ex}_{\ell}$ ($\ell=1,2,3)$ are eigenstates with large overlap with  $|S_{\ell}\rangle$, denoted by $|S^{e}_{\ell}\rangle$, and the spectrum lines of $|S^{e}_{\ell}\rangle$ intersect with other spectrum lines of the other eigenstates as increasing $\mu_d$. 

Here, we compare the lowest energy spectral line (the orange line in Fig.~\ref{Fig8} (a)) of $|S^{e}_{1}\rangle$ 
with the first-order corrected energy of $|S_{\ell}\rangle$ denoted by $E^{(1)}_{1}$. There is no difference between them as shown in Fig~~\ref{Fig8} (b). 
Trivially, the energy splitting is well captured by the first-order degenrate perturbation theory. 
We also calculated some overlaps in Fig.~\ref{Fig8} (c) and (d). 
Even for small finite $\mu_d$, the overlap between $|S^{e}_{\ell}\rangle$ and $|S_{\ell}\rangle$ is large [See Fig.~\ref{Fig8} (c)] and also the overlap between $|S^{e}_{1}\rangle$ and the first-order corrected state $|S^{(1)}_{1}\rangle$ obtained by Eq.~(\ref{1stPstate}) is large although some accidental weak breakdown occurs due to the level crossing to the other eigenstates. However, the deviation is small since more than 95$\%$ overlap appears. 
These results imply that $|S^{e}_{\ell}\rangle$ almost inherits properties of the CLS many-body states $|S_{\ell}\rangle$ under small $\mu_d$.

We further observe the behavior of the EE of $|S^{e}_{1}\rangle$ and $|S^{(1)}_1\rangle$ obtained from Eq.~(\ref{1stPstate}), where the EE is obtained from half of the system.  
The result is plotted in Fig.~\ref{Fig8} (e). 
The EE of $|S^{(1)}_{1}\rangle$ keeps low value with a slight increase 
coming from the mixing of the other eigenstates.  
However, this increase is much small compared to the order of EE of other eigenstates around $E=0$, $\sim \mathcal{O}(10^{-1})$ as shown in Fig.~\ref{Fig7} (c). 
The EE of $|S^e_{1}\rangle$ also keeps low-valued within our target regime of $\mu_d$ as a whole. 
There are two accidental peaks in the EE of $|S^e_{1}\rangle$, where the accidental deviation from $|S_{1}\rangle$ occurs in Fig.~\ref{Fig8} (c) and (d). These increases are small compared to the order of EE of other eigenstates around $E=0$, $\sim \mathcal{O}(10^{-1})$ as shown in Fig.~\ref{Fig7} (c). 
Interestingly, even if the EE of $|S^{(e)}_1\rangle$ increases near the intersection, after the intersection it takes low-valued EE close to the EE of the first-order perturbation result when the state of $|S^{e}_1\rangle$ is isolated again. Accordingly, these numerical results imply that with small $\mu_d$, the state $|S^{e}_{\ell}\rangle$ remains low-entangled even for an accidental hybridization, and its properties are well captured by the first-order degenerate perturbation theory.

\section*{Appendix D: Analytical solution of SRP in two particle system}

For $N_p$ particle system with $L$, we show the analytical form of the SRP [Eq.~(\ref{SRP})] for the initial state $|S^{D}{(N_p,L)}\rangle$ with $N_D$. 
Directly from the matrix element of $\langle S_{\ell}|\hat{V}_{p}|S_{\ell}\rangle $ ($\ell=1,\cdots, N_D$), the return probability is given by
\begin{eqnarray}
\langle S^{D}|e^{-i\hat{H}_{tot}t}|S^{D}\rangle
\sim \frac{1}{N_D}\biggl|\sum^{N_D}_{\ell=1}e^{-i\epsilon_{\ell}t} \biggr|^{2},
\end{eqnarray}
where we use an approximation, $e^{-i\hat{H}_{tot}t}|S_{\ell}\rangle \sim e^{-i\epsilon_{\ell}t}|S_{\ell}\rangle$. 
For $N_p=2$ system with $L=3$, $N_D=3$ and $\epsilon_{\ell}=(-7.5+4\ell)\mu_d$ ($\ell=1,2,3$) are given.
Roughly speaking, the SRP oscillates with the two frequency, $\delta \epsilon_{21}/\hbar$, $\delta \epsilon_{31}/\hbar$, where $\delta \epsilon_{kj}=\epsilon_{k}-\epsilon_{j}$. This is consistent to the numerical results in Fig.~\ref{Fig7} (e). The analytical solution of the SRP for $L=5$ and $N_p=4$ can be directly calculated in the same way.  
The analytical solution is given by $N_D=5$ and $\epsilon_{\ell}=(-10+4\ell)\mu_d$ ($\ell=1,2,3,4,5$). 
This solution is also plotted in Fig.~\ref{Fig7} (a) and (b). 

\section*{Appendix E: Degenerate perturbation theory for four particle system}
In this appendix, we show the results of the degenerate perturbation theory for $L=5$ and $N_p=4$ system in the same way as in Appendix C. 
For this system, without $\hat{V}_p$, there are five CLS many-body states of Eq.~(\ref{SD_ND=5}) with zero energy.
Under the perturbation $\hat{V}_{p}$, the degeneracy of the CLS many-body states is lifted and these CLS states are slightly corrected. 
Here we focus on the regime $\mu_d \in [0:0.015]$. 
Figure~\ref{Fig9} (a) is the flow of the exact spectrum around $E=0$ with the increase of $\mu_d$ by exact diagonalization. 
The spectral lines extending from $E = 0$ at $\mu_d=0$ denoted by $E^{ex}_{\ell}$ ($\ell=1,2,3,4,5)$ are the exact eigenstates denoted by $|S^{e}_{\ell}\rangle$ with large overlap with $|S_{\ell}\rangle$. 
The degeneracy is clearly lifted for $\mu_d \neq 0$. 
The spectrum lines of $|S^{e}_{\ell}\rangle$ intersect with many other spectrum lines of the other eigenstates. 
We compare the energy spectral line with the larger overlap for $|S_{5}\rangle$ (the orange line in Fig.~\ref{Fig9} (a)) of $|S^{e}_{5}\rangle$ 
with the first-order corrected energy of $|S_{5}\rangle$ denoted by $E^{(1)}_{5}$. 
They show a good agreement as shown in Fig~~\ref{Fig9} (b). 
The energy splitting is well captured by the first-order degenrate perturbation theory as in the two-particle case. 
We also calculated some overlap in Fig.~\ref{Fig9} (c) and (d). 
As a whole, even for small finite $\mu_d$, the overlap between $|S^{e}_{\ell}\rangle$ and $|S_{\ell}\rangle$ is large. 
However, since there are many level crossing of the other energy eigenstates compared to the two-particle case, more accidental deviations occur [See Fig.~\ref{Fig9} (c)]. 
The overlap between $|S^{e}_{5}\rangle$ and the first-order perturbed state $|S^{(1)}_{5}\rangle$ obtained by Eq.~(\ref{1stPstate}) shows the same tendency, as a whole; the large overlap is kept although some accidental breakdowns occur due to some specific level crossings to the other eigenstates. 
The deviation however is very small since more than 99$\%$ overlap appears.  
As in the two-particle case, these results also imply that $|S^{e}_{\ell}\rangle$ mostly inherits properties of the CLS many-body states $|S_{\ell}\rangle$ under small $\mu_d$.

We finally observe the behavior of the EEs of $|S^{e}_{5}\rangle$ and $|S^{(1)}_5\rangle$ obtained from Eq.~(\ref{1stPstate}). 
The result is plotted in Fig.~\ref{Fig9} (d). The EE of $|S^{(1)}_{5}\rangle$ keeps low value with a slight increase 
coming from the mixing to the other eigenstates. 
However, this increase is much smaller compared to the order of EE of other eigenstates around $E=0$, $\sim \mathcal{O}(10^{-1})$ as shown in Fig.~\ref{Fig3} (b).
The EE of $|S^{e}_{5}\rangle$ also shows several peaks. 
The several accidental peaks of EE originate from the hybridization of other eigenstates crossing to the state $|S^{e}_{5}\rangle$ at the $\mu_d$ around the peaks as shown in Fig.~\ref{Fig8} (a). These increases from the peaks are also small compared to the order of EE of other eigenstates around $E=0$, $\sim \mathcal{O}(10^{-1})$ as shown in Fig.~\ref{Fig3} (b). Hence, as a whole, the EE of $|S^e_{5}\rangle$ keeps low-valued within our target regime of $\mu_d$. 
Actually, at $\mu_d=0.01$, where the numerical results are presented in Fig.~\ref{Fig4}, the EE remains low-valued and much close to the result of EE obtained by the first-order corrected states $|S^{(1)}_5\rangle$.  
Accordingly, these numerical results imply that 
with small $\mu_d$, the states $|S^{e}_{\ell}\rangle$ are low-entangled even for an accidental hybridization, and its properties are well captured by the first-order degenerate perturbation theory.


\begin{thebibliography}{99}

\bibitem{AL50}
A. Lagendijk, B. v. Tiggelen, and D. Wiersma, Physics
Today {\bf 62}, 24 (2009).

\bibitem{Deutsch}
J. M. Deutsch, Phys. Rev. A {\bf 43}, 2046 (1991).

\bibitem{D'Alessio}
L. D'Alessio, Y. Kafri, A. Polkovnikov, and
M. Rigol, Advances in Physics {\bf 65}, 239 (2016)

\bibitem{Srednicki}
M. Srednicki, Phys. Rev. E {\bf 50}, 888 (1994).

\bibitem{Rigol}
M. Rigol, V. Dunjko, V. Yurovsky, and M. Olshanii,
Phys. Rev. Lett. {\bf 98}, 050405 (2007).

\bibitem{Garrison}
J. R. Garrison and T. Grover, Phys. Rev. X {\bf 8}, 021026
(2018).

\bibitem{Kim}
H. Kim, T. N. Ikeda, and D. A. Huse, Phys. Rev. E {\bf 90},
052105 (2014).

\bibitem{Santos}
L. F. Santos and M. Rigol, Phys. Rev. E {\bf 81}, 036206
(2010).

\bibitem{Choi2015}
J.-y. Choi, S. Hild, J. Zeiher, P. Schau{\ss}, A. RubioAbadal, T. Yefsah, V. Khemani, D. A. Huse, I. Bloch,
and C. Gross, Science {\bf 352}, 1547 (2016).

\bibitem{Schreiber}
M. Schreiber, S. S. Hodgman, P. Bordia, H. P. L{\"u}schen,
M. H. Fischer, R. Vosk, E. Altman, U. Schneider, and
I. Bloch, Science {\bf 349}, 842 (2015).

\bibitem{Lukin2019}
A. Lukin, M. Rispoli, R. Schittko, M. E. Tai, A. M. Kaufman,
S. Choi, V. Khemani, J. Leonard, and M. Greiner,
Science (New York, N.Y.) {\bf 364}, 256-260 (2019).

\bibitem{Nandkishore}
R. Nandkishore and D. A. Huse, Annual Review
of Condensed Matter Physics {\bf 6}, 15 (2015).

\bibitem{serbyn2020}
M. Serbyn, D. A. Abanin, and Z. Papi\'{c}, ``Quantum
many-body scars and weak breaking of ergodicity,"
(2020), arXiv:2011.09486 [quant-ph].

\bibitem{Abanin2019}
D. A. Abanin, E. Altman, I. Bloch, and M. Serbyn, Rev.
Mod. Phys. {\bf 91}, 021001 (2019).

\bibitem{Imbrie2017}
J. Z. Imbrie, V. Ros, and A. Scardicchio, Annalen der
Physik {\bf 529}, 1600278 (2017).

\bibitem{Bernien}
H. Bernien, S. Schwartz, A. Keesling, H. Levine, A. Omran,
H. Pichler, S. Choi, A. S. Zibrov, M. Endres,
M. Greiner, V. Vuleti\'{c}, and M. D. Lukin, Nature {\bf 551},
579 (2017).

\bibitem{Ebadi2020}
S. Ebadi, T. T. Wang, H. Levine, A. Keesling, G. Semeghini,
A. Omran, D. Bluvstein, R. Samajdar, H. Pichler,
W. W. Ho, S. Choi, S. Sachdev, M. Greiner,
V. Vuletic, and M. D. Lukin, ``Quantum phases of matter
on a 256-atom programmable quantum simulator,"
(2020), arXiv:2012.12281 [quant-ph].

\bibitem{Bluvstein}
D. Bluvstein, A. Omran, H. Levine, A. Keesling, G. Semeghini,
S. Ebadi, T. T. Wang, A. A. Michailidis,
N. Maskara, W. W. Ho, S. Choi, M. Serbyn, M. Greiner,
V. Vuletic, and M. D. Lukin, ``Controlling many-body
dynamics with driven quantum scars in rydberg atom
arrays," (2020), arXiv:2012.12276 [quant-ph].

\bibitem{Turner}
C. J. Turner, A. A. Michailidis, D. A. Abanin, M. Serbyn,
Papi\'{c}, and Z. , Nature Physics {\bf 14}, 745 (2018).

\bibitem{Turner2}
C. J. Turner, A. A. Michailidis, D. A. Abanin, M. Serbyn,
and Z. Papi\'{c}, Phys. Rev. B {\bf 98}, 155134 (2018).

\bibitem{Choi}
S. Choi, C. J. Turner, H. Pichler, W. W. Ho, A. A.
Michailidis, Z. Papi\'{c}, M. Serbyn, M. D. Lukin, and D. A.
Abanin, Phys. Rev. Lett. {\bf 122}, 220603 (2019).

\bibitem{Ho2019}
W. W. Ho, S. Choi, H. Pichler, and M. D. Lukin, Phys.
Rev. Lett. {\bf 122}, 040603 (2019).

\bibitem{Moudgalya1}
S. Moudgalya, N. Regnault, and B. A. Bernevig, Phys. Rev. B {\bf 98}, 235156 (2018).

\bibitem{Moudgalya2}
S. Moudgalya, S. Rachel, B. A. Bernevig, and N. Regnault,
Phys. Rev. B {\bf 98}, 235155 (2018).

\bibitem{Schecter2019}
M. Schecter and T. Iadecola, Phys. Rev. Lett. {\bf 123},
147201 (2019).

\bibitem{Iadecola2019}
T. Iadecola, M. Schecter, and S. Xu, Phys. Rev. B {\bf 100},
184312 (2019).

\bibitem{Lin2019}
C.-J. Lin and O. I. Motrunich, Phys. Rev. Lett. {\bf 122},
173401 (2019).

\bibitem{Iadecola2020}
T. Iadecola and M. Schecter, Phys. Rev. B {\bf 101}, 024306
(2020).

\bibitem{McClarty}
P. A. McClarty, M. Haque, A. Sen, and J. Richter, Phys.
Rev. B {\bf 102}, 224303 (2020).

\bibitem{Wildeboer}
J. Wildeboer, A. Seidel, N. S. Srivatsa, A. E. B. Nielsen,
and O. Erten, ``Topological quantum many-body scars in
quantum dimer models on the kagome lattice," (2020),
arXiv:2009.00022 [cond-mat.str-el].

\bibitem{Srivatsa}
N. S. Srivatsa, J. Wildeboer, A. Seidel, and A. E. B.
Nielsen, Phys. Rev. B {\bf 102}, 235106 (2020).

\bibitem{Shibata}
N. Shibata, N. Yoshioka, and H. Katsura, Phys. Rev.
Lett. {\bf 124}, 180604 (2020).

\bibitem{Lee2020}
K. Lee, R. Melendrez, A. Pal, and H. J. Changlani, Phys.
Rev. B {\bf 101}, 241111 (2020).

\bibitem{Huang}
K. Huang, Y.Wang, and X. Li,``Stability of 2d quantum
many-body scar states against random disorder," (2021),
arXiv:2102.08241 [cond-mat.dis-nn].

\bibitem{Langlett}
C. M. Langlett and S. Xu, ``Hilbert space fragmentation
and exact scars of generalized fredkin spin chains,"
(2021), arXiv:2102.06111 [cond-mat.str-el].

\bibitem{Zhao2021}
H. Zhao, A. Smith, F. Mintert, and J. Knolle,
``Orthogonal quantum many-body scars," (2021),
arXiv:2102.07672 [cond-mat.stat-mech].

\bibitem{Ok}
S. Ok, K. Choo, C. Mudry, C. Castelnovo, C. Chamon,
and T. Neupert, Phys. Rev. Research {\bf 1}, 033144 (2019).

\bibitem{Moudgalya2020}
S. Moudgalya, B. A. Bernevig, and N. Regnault, Phys.
Rev. B {\bf 102}, 195150 (2020).

\bibitem{Jeyaretnam}
J. Jeyaretnam, J. Richter, and A. Pal, ``Quantum scars
and bulk coherence in a symmetry-protected topological
phase," (2021), arXiv:2103.15880 [cond-mat.stat-mech].

\bibitem{Hudomal}
A. Hudomal, I. Vasi\'{c}, N. Regnault, and Z. Papi\'{c}, Communications Physics {\bf 3}, 99 (2020).

\bibitem{Sinha}
S. Sinha and S. Sinha, Phys. Rev. Lett. {\bf 125}, 134101
(2020).

\bibitem{Zhao2020}
H. Zhao, J. Vovrosh, F. Mintert, and J. Knolle, Phys.
Rev. Lett. {\bf 124}, 160604 (2020).

\bibitem{Surace}
F. M. Surace, P. P. Mazza, G. Giudici, A. Lerose,
A. Gambassi, and M. Dalmonte, Phys. Rev. X {\bf 10}, 021041 (2020).

\bibitem{Banerjee}
D. Banerjee and A. Sen, (2020), arXiv:2012.08540 [condmat.
str-el].

\bibitem{Yang}
C. N. Yang, Phys. Rev. Lett. {\bf 63}, 2144 (1989).

\bibitem{Zhang}
S. Zhang, Phys. Rev. Lett. {\bf 65}, 120 (1990).

\bibitem{Vafek2017}
O. Vafek, N. Regnault, and B. A. Bernevig, SciPost
Phys. {\bf 3}, 043 (2017).

\bibitem{Mark}
D. K. Mark and O. I. Motrunich, Phys. Rev. B {\bf 102},
075132 (2020).

\bibitem{DeTomasi2019}
G. De Tomasi, D. Hetterich, P. Sala, and F. Pollmann, Phys. Rev. B {\bf 100}, 214313 (2019).

\bibitem{KMH2020}
Y. Kuno, T. Mizoguchi, and Y. Hatsugai, Phys. Rev. B
{\bf 102}, 241115 (R) (2020).

\bibitem{Scherg}
S. Scherg, T. Kohlert, P. Sala, F. Pollmann, B. H.
M., I. Bloch, and M. Aidelsburger, ``Observing nonergodicity
due to kinetic constraints in tilted fermihubbard
chains," (2020), arXiv:2010.12965 [condmat.
quant-gas].

\bibitem{Khemani2020}
V. Khemani, M. Hermele, and R. Nandkishore, Phys.
Rev. B {\bf 101}, 174204 (2020).

\bibitem{Lin}
C.-J. Lin, A. Chandran, and O. I. Motrunich, Phys. Rev.
Research {\bf 2}, 033044 (2020).

\bibitem{remark1}
For $\mu_0<0$, the flat band in the single particle spectrum of $\hat{H}_{0}$ is the second band while for $\mu_0>0$, the flat band is lowest.

\bibitem{Maruyama}
Y. Hatsugai and I. Maruyama, EPL (Europhysics Letters)
{\bf 95}, 20003 (2011).

\bibitem{Mizoguchi}
T. Mizoguchi and Y. Hatsugai, EPL (Europhysics Letters)
{\bf 127}, 47001 (2019).

\bibitem{Mizoguchi2021}
T. Mizoguchi, Y. Kuno, and Y. Hatsugai, ``Flat band,
spin-1 dirac cone, and hofstadter diagram in fermionic
square kagome model," (2021), arXiv:2103.03489 [condmat.
mes-hall].

\bibitem{Hatsugai2021}
Y. Hatsugai, Annals of Physics, 168453 (2021).

\bibitem{Meinert}
F. Meinert, M. J. Mark, E. Kirilov, K. Lauber, P. Weinmann,
M. Grobner, A. J. Daley, and H.-C. Nagerl, Science
(New York, N.Y.) {\bf 344}, 1259?1262 (2014).

\bibitem{Hart}
O. Hart, G. De Tomasi, and C. Castelnovo, Phys. Rev. Research {\bf 2}, 043267 (2020).

\bibitem{t_j_case2}
We exclude a random distribution $t_j$ since, if the randomness
is large, some localizations can appear.

\bibitem{Zhang2015}
T. Zhang and G.-B. Jo, Scientific Reports {\bf 5}, 16044
(2015).

\bibitem{tilt_scale}
The system size dependence of the energy scale of $\hat{V}_p$ is $\mathcal{O}(4\mu_d L)$. We assume that $\mu_d$ is set to be very small, where the energy scale  $\mathcal{O}(4\mu_d L)$ is smaller than other model parameters.

\bibitem{Schulz}
M. Schulz, C. A. Hooley, R. Moessner, and F. Pollmann,
Phys. Rev. Lett. {\bf 122}, 040606 (2019).

\bibitem{Nieuwenburg}
E. van Nieuwenburg, Y. Baum, and
G. Refael, Proceedings of the National
Academy of Sciences {\bf 116}, 9269 (2019).

\bibitem{Orito}
T. Orito, Y. Kuno, and I. Ichinose, Phys. Rev. B {\bf 100},
214202 (2019).

\bibitem{Taylor}
S. R. Taylor, M. Schulz, F. Pollmann, and R. Moessner,
Phys. Rev. B {\bf 102}, 054206 (2020).

\bibitem{Quspin}
For all numerical simulations, we employed the Quspin
solver: P. Weinberg and M. Bukov, SciPost Phys. {\bf 7}, 20
(2019); {\bf 2}, 003 (2017).

\bibitem{Pal}
A. Pal and D. A. Huse, Phys. Rev. B {\bf 82}, 174411 (2010).

\bibitem{Page_curve}
Strictly speaking, the small deviation from the EE of volume law $S^{Max}_e$ comes from the presence of Page curve.

\bibitem{exp}
Experimentally, it is possible to prepare the initial state: 
one set the noninteracting system ${\hat H}_0$ with $\mu_d>0$ and the particle number fixed.

\bibitem{Khemani}
V. Khemani, C. R. Laumann, and A. Chandran, Phys.
Rev. B {\bf 99}, 161101 (2019).

\end{thebibliography}

\end{document}